\documentclass[pra, twocolumn, amscd, amsmath, amssymb,letterpaper,showpacs,biblatex,superscriptaddress]{revtex4-1}

\usepackage{graphicx}
\usepackage{bm}
\usepackage{amsfonts}
\usepackage{eurosym}
\usepackage{amsmath, amsthm, amssymb}
\usepackage{tabularx,colortbl}
\usepackage{dcolumn}   
\usepackage{array}
\usepackage{amscd}
\usepackage{dsfont}
\usepackage{color}
\usepackage{verbatim}
\usepackage{subfigure}
\usepackage{bbold}

\newcommand{\beq}{\begin{equation}}
\newcommand{\eeq}{\end{equation}}
\newcommand{\beqa}{\begin{eqnarray}}
\newcommand{\eeqa}{\end{eqnarray}}

\newcommand{\Ub}{\mathbb{U}}
\newcommand{\ket}[1]{|#1\rangle}
\newcommand{\eq}{\overset{\text{l.u.}}{=}}

\begin{document}


\title{Mediated gates between spin qubits}

\author{Jianjia Fei}
\email{jfei@wisc.edu}
\affiliation{Department of Physics, University of Wisconsin-Madison, Madison, Wisconsin 53706, USA}
\author{Dong Zhou}
\affiliation{Department of Physics, Yale University, New Haven, Connecticut 06520, USA}
\author{Yun-Pil Shim}
\affiliation{Department of Physics, University of Wisconsin-Madison, Madison, Wisconsin 53706, USA}
\author{Sangchul Oh}
\affiliation{Department of Physics, University at Buffalo, State University of New York, Buffalo, New York 14260, USA}
\author{Xuedong Hu}
\affiliation{Department of Physics, University at Buffalo, State University of New York, Buffalo, New York 14260, USA}
\author{Mark Friesen}
\affiliation{Department of Physics, University of Wisconsin-Madison, Madison, Wisconsin 53706, USA}

\date{\today}

\begin{abstract}
In a typical quantum circuit, nonlocal quantum gates are applied to nonproximal qubits. 
If the underlying physical interactions are short-range (\emph{e.g.}, exchange interactions between spins),  intermediate \textsc{swap} operations must be introduced, thus increasing the circuit depth.
Here we develop a class of ``mediated" gates for spin qubits, which act on nonproximal spins via intermediate ancilla qubits.
At the end of the operation, the ancillae return to their initial states.
We show how these mediated gates can be used (1) to generate arbitrary quantum states and (2) to construct arbitrary quantum gates.
We provide some explicit examples of circuits that generate common states [\emph{e.g.}, Bell, Greenberger-Horne-Zeilinger (GHZ), $W$, and cluster states] and gates (\emph{e.g.}, $\sqrt{\textsc{swap}}$, \textsc{swap}, \textsc{cnot}, and Toffoli gates).
We show that the depths of these circuits are often shorter than those of conventional $\textsc{swap}$-based circuits. 
We also provide an explicit experimental proposal for implementing a mediated gate in a triple-quantum-dot system.
\end{abstract}

\maketitle

\section{Introduction}
Quantum dot spin qubits are promising candidates for quantum computing because of their long decoherence times and their potential to leverage existing semiconductor technologies~\cite{Hanson07, Morton11}. 
The exchange coupling is a desirable tool for mediating interactions between spin qubits, because it can be controlled electrostatically and it is typically very fast~\cite{Petta05}.
In combination with arbitrary single-qubit operations, the exchange coupling enables universal quantum computation~\cite{Loss98}.
When logical qubits consist of two~\cite{Levy02} or three~\cite{DiVincenzo00} physical qubits in a decoherence-free subsystem, the exchange coupling alone is universal for quantum computation.
On the other hand, the intrinsic short-range nature of the exchange coupling (typically tens of nanometers) imposes strong constraints on the physical architecture of the spin qubits. 
These constraints present a significant challenge to scalability during quantum error correction, particularly for linear qubit architectures, which are typical for quantum dot spin qubits~\cite{Svore05}.  
Indeed, large-scale quantum computing is challenging in any qubit implementation, and the complexity of a given quantum circuit could, to a large extent, determine its success. 

In a number of qubit systems, such as nuclear magnetic resonance (NMR), the physical interactions may be constant or ``always on." 
This is not necessarily a disadvantage.
For example, it has been shown that simultaneous, multiqubit couplings can be used to enable quantum state transfer~\cite{Khaneja02, Oh11}, and other rudimentary quantum gates~\cite{Yung04}.  
Similar considerations apply to quantum dot spin systems with Heisenberg couplings~\cite{Benjamin03}.  
Quantum dots provide unique opportunities for controlling the nature of the interactions.  
For example, simultaneous, multiqubit couplings could provide a potential route for enhancing the effective range of the coupling, in analogy with the Ruderman-Kittel-Kasuya-Yosida (RKKY) interaction~\cite{Ruderman54}.
When these couplings are arranged into nontrivial topologies, such as rings, a rich spectrum of quantum gates emerges~\cite{Mizel04, Scarola05, Hsieh12}.
However, even simple topologies, like those considered here, can produce entangling gates that differ from the existing two-qubit gates in spin qubits~\cite{Shulman12, Brunner11, Nowack11}.

\begin{figure}[h]
\centering
\includegraphics*[width=.5\linewidth]{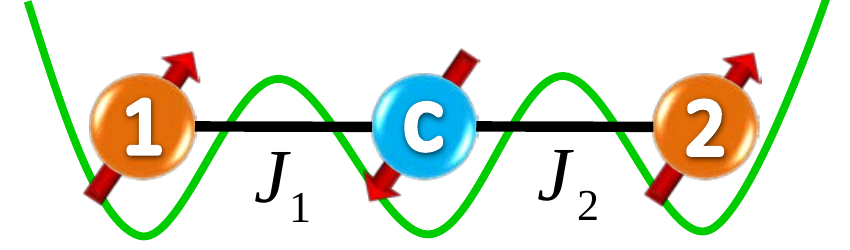}
\caption{(Color online) 
A linear triple-quantum-dot geometry with three electrons. 
Mediated gates can be achieved between qubit 1 and qubit 2 by applying simultaneous exchange couplings, with $J_1=J_2$. 
At the end of the operation, the ancilla qubit $c$ is restored to its initial state.}
\label{fig:1+2}
\end{figure}

In this paper, we show how to control such simultaneous, multiqubit couplings.
The result is a class of ``mediated" quantum gates.
We focus primarily on the three-qubit geometry shown in Fig.~\ref{fig:1+2}, due to recent experimental progress on triple quantum dots~\cite{Gaudreau06,Laird10,Gaudreau11}.
In this arrangement, the mediated gate acts on the nonproximal qubits 1 and 2, leaving the ancilla or central qubit $c$ unaffected, at the end of the operation.
We characterize this well-defined gate operation, $\Ub_2$, and show how arbitrary two-qubit states and gates can be generated using $\Ub_2$ as the sole entangling resource.
We also compare the circuit depth of these mediated gate protocols to more conventional \textsc{swap}-based protocols.
Finally, we explain how long-range mediated gates can be attained by replacing qubit $c$ with a spin bus.

\section{Two-Qubit Mediated Gate, $\mathbb{U}_2$}
\label{sec:U2}
\subsection{Mediated gate, $\mathbb{U}_2$}
We begin by characterizing the mediated gate $\Ub_2$.
The effective spin Hamiltonian for the quantum dot geometry in Fig.~\ref{fig:1+2} derives from the exchange interaction and takes the form of nearest-neighbor Heisenberg couplings~\cite{Loss98}, given by
\begin{equation}
H=J_1\mathbf{s}_1\cdot\mathbf{s}_c+J_2\mathbf{s}_2\cdot\mathbf{s}_c,
\end{equation} 
where $\mathbf{s}_j$ are spin operators.
In typical experiments, the coupling constants $J_1$ and $J_2$ are controlled by detuning the local electrostatic potentials in a given dot~\cite{Petta05}.
$J_1$ and $J_2$ can usually be varied independently as a function of time~\cite{Gaudreau06,Laird10}.
For mediated gates, however, we assume that both couplings are turned on and off simultaneously.

The Hamiltonian $H$ induces three-spin dynamics according to the time evolution operator $U(t)=e^{-iHt}$,  where we set $\hbar=1$.
However, the mediated gate we seek has the special form $U=\Ub_2 \otimes I$, where $\Ub_2$ acts on qubits 1 and 2, and the identity operator $I$ acts on the ancilla qubit $c$.
In Appendix~A, we prove that only one nontrivial mediated gate exists for the geometry in Fig.~\ref{fig:1+2}, corresponding to the unique parameter combination $J_1=J_2=J$ and the special evolution period $T_g=4\pi/3J$ (with periodic recurrences~\cite{Friesen07}).
The gate is robust against control errors, similarly to conventional two-qubit exchange gates.
For example, if $J_2=J_1(1+\delta)$ results in the gate $U(\delta)$, where $U(0)$ is the desired gate, and if the fidelity is defined as~\cite{Hill07} $F=|\text{Tr} [U(\delta)^\dagger U(0)]|/\text{Tr} [U(0)^\dagger U(0)]$,
then we obtain a quadratic error in the fidelity: $1-F\simeq 0.97\delta^2$ when $\delta\le 0.4$.

\begin{figure}[h]
\centering
\includegraphics*[width=.8\linewidth]{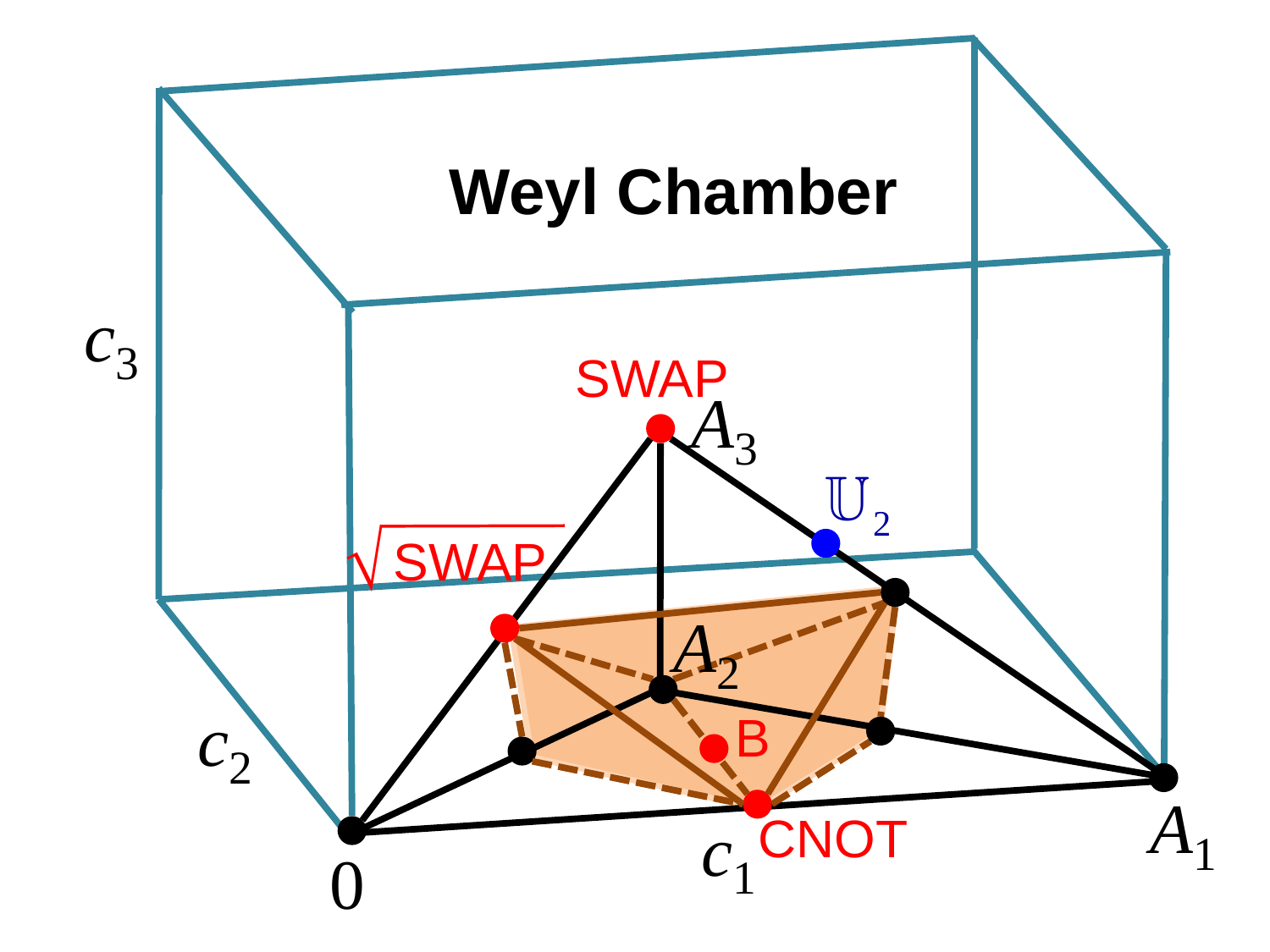}
\caption{(Color online) 
A geometric representation of two-qubit SU(4) gate operations, with axes $c_1, c_2$, and $c_3$ defined in Eq.~(\ref{eq:Cartan}).
The Weyl chamber corresponds to the tetrahedron $0$-$A_1$-$A_2$-$A_3$,
while the gates known as perfect entanglers lie inside the shaded region~\cite{JZhang03pra}. 
The coordinates for other special gates are given by $\textsc{b}=(\frac{\pi}{2},\frac{\pi}{4},0)$, $\textsc{cnot}=(\frac{\pi}{2},0,0)$, $\textsc{swap}=(\frac{\pi}{2},\frac{\pi}{2},\frac{\pi}{2})$, and $\sqrt{\textsc{swap}}=(\frac{\pi}{4},\frac{\pi}{4},\frac{\pi}{4})$.}
\label{fig:tetra}
\end{figure}

Any unitary two-qubit operator $U_2\in \text{SU}(4)$, including $\Ub_2$, can be expressed in the form of a Cartan decomposition, given by~\cite{Kraus01, JZhang03pra}
\begin{equation}
\label{eq:Cartan}
U_2\eq e^{\frac{i}{2}(c_1\sigma_x\otimes\sigma_x+c_2\sigma_y\otimes\sigma_y+c_3\sigma_z\otimes\sigma_z)},
\end{equation}
where $\sigma_x, \sigma_y, \sigma_z$ are the Pauli matrices, and $\mathbf{s}={\bm \sigma}/2$ in spinor notation.
Here, the relation $\eq$ means ``equal, up to local unitary gates," where the latter may be applied before and/or after the nonlocal operator.  
The decomposition is unique when the parameters $(c_1, c_2, c_3)$ are restricted to the tetrahedron $\pi-c_2\geq c_1\geq c_2\geq c_3\geq 0$, known as the Weyl chamber. (Note that special considerations apply to the base of the tetrahedron~\cite{JZhang03pra}.)
There is a one-to-one mapping between the Weyl chamber and the Makhlin invariants~\cite{Makhlin02}, which provides an alternative representation of the nonlocal properties of $U_2\in \text{SU}(4)$ (except on the bottom surface of the chamber).
The Cartan decomposition for our two-qubit mediated gate is given by $(c_1,c_2,c_3)=(2,1,1)(\pi/3)$ and it has the explicit form (see Appendix~A for details)
\begin{widetext}
\begin{equation}
\mathbb{U}_{2} =
-\left( {\begin{array}{cccc}
 \frac{1}{2}(1+i\sqrt{3}) & 0 & 0 & 0  \\
 0 & \frac{1}{4}(-1+i\sqrt{3}) & \frac{1}{4}(3+i\sqrt{3})  & 0  \\
 0 & \frac{1}{4}(3+i\sqrt{3})  & \frac{1}{4}(-1+i\sqrt{3}) & 0  \\
 0 & 0 & 0 & \frac{1}{2}(1+i\sqrt{3})\\
 \end{array} } \right).
\label{eq:U2def}
\end{equation}
\end{widetext}
The position of $\Ub_2$ in the Weyl chamber is shown in Fig.~\ref{fig:tetra}, along with several other common two-qubit gates.

The gating capabilities of $U_2\in \text{SU}(4)$ derive from its entangling properties, which can be characterized in part by its position in the Weyl chamber.
The operators known as ``perfect entanglers" lie inside a polyhedron, which fills half of the chamber~\cite{JZhang03pra}, as shown in Fig.~\ref{fig:tetra}.
Combined with local unitaries, a perfect entangler can generate a maximally entangled state from a separable state. 
For example, if we quantify two-qubit entanglement in terms of the ``concurrence" measure $C$~\cite{Hill97, Wootters98}, then a separable state exhibits no entanglement, with $C=0$, while a highly nonlocal state like the singlet Bell state $\vert\Psi^-\rangle=\frac{1}{\sqrt{2}}(\vert 01\rangle-\vert 10\rangle)$ exhibits maximal entanglement, with $C=1$.
Thus, for some initial two-qubit state with $C=0$, one application of a perfect entangler produces a state with $C=1$.
The standard \textsc{cnot} gate is known to be a perfect entangler~\cite{JZhang03pra}, as indicated in Fig.~\ref{fig:tetra}. 
However, \textsc{cnot} does not arise naturally from the exchange interaction between spin qubits; it must be constructed from more basic gates~\cite{Loss98}.
In contrast, the mediated gate $\Ub_2$ does arise naturally in many-body spin systems as we have  shown; however, its location in the Weyl chamber indicates that it is not a perfect entangler.
Using the methods of~\cite{Kraus01}, we find that $\Ub_2$ achieves a maximum concurrence of $C_\text{max}=\sqrt{3}/2 < 1$, when acting on a separable state. 

A universal quantum processor must be able to generate arbitrary entangled states or implement arbitrary quantum circuits. 
For example, \textsc{cnot} gates combined with single-qubit unitaries are known to be universal~\cite{Barenco95, Nielsen&Chuang, Vidal04}. 
Any two-qubit entangling gate $U_2$ can replace \textsc{cnot} in this scheme~\cite{Bremner02}, although the entangling capabilities of the gate will affect the overall circuit depth. 
In the remainder of this section, we explore methods for generating arbitrary entangled states and entangling gates using the mediated gate $\Ub_2$, and we determine the circuit depth of such protocols.

\subsection{Generation of arbitrary states}
\label{sec:U2_state}
Our goal here is to construct arbitrary, two-qubit, entangled pure states between qubit 1 and qubit 2 in the geometry shown in Fig.~\ref{fig:1+2}, using the mediated gate $\Ub_2$ as our nonlocal entangling resource. 
Allowing for local operations and classical communication (LOCC), it is possible to transform maximally entangled states, such as Bell states, into arbitrary pure states~\cite{Nielsen99}.
We therefore focus on using $\Ub_2$ to generate Bell states. 
For simplicity, we ignore global phase factors throughout this paper.

The strategy we adopt is to apply $\mathbb{U}_2$ repeatedly, assisted by single-qubit unitary rotations $U_1$, as needed:
\begin{equation}
\label{eq:rep}
\ket{\psi} = (U_1\otimes U_1)[\mathbb{U}_2  (U_1\otimes U_1)]^n \ket{00} .
\end{equation}
Note that each application of $U_1$ here represents an arbitrary rotation, and that, in general, the rotations can all be different. 

We would like to be able to compare the speed or efficiency of disparate gating protocols, particularly between mediated and conventional gates.
The most convenient measure of this efficiency is the ``circuit depth," which we define here as the total number of exchange gates.
For example, in Eq.~(\ref{eq:rep}), the circuit depth is equal to $n$.
For conventional quantum dot circuits, there may be cases where it is possible to implement gates between different pairs of qubits simultaneously, due to physical separation.
We define the circuit depth of such parallel gates to be 1, since they occur simultaneously.
On the other hand, conventional circuits typically require intermediate \textsc{swap} gates to be applied sequentially when the qubits are nonproximal, causing the circuit depth to increase by 1 with each  \textsc{swap} application.
This notion of circuit depth plays an important role in the gate times and fidelities of quantum circuits, and we speak of circuit depth throughout the following discussion.
To conclude, we note that since $\mathbb{U}_2$ is not a perfect entangler, the value of $n$ in Eq.~(\ref{eq:rep}) must be greater than 1 when we generate a Bell state. 

\begin{figure}[t]
\centering
\includegraphics[width=1.0\linewidth]{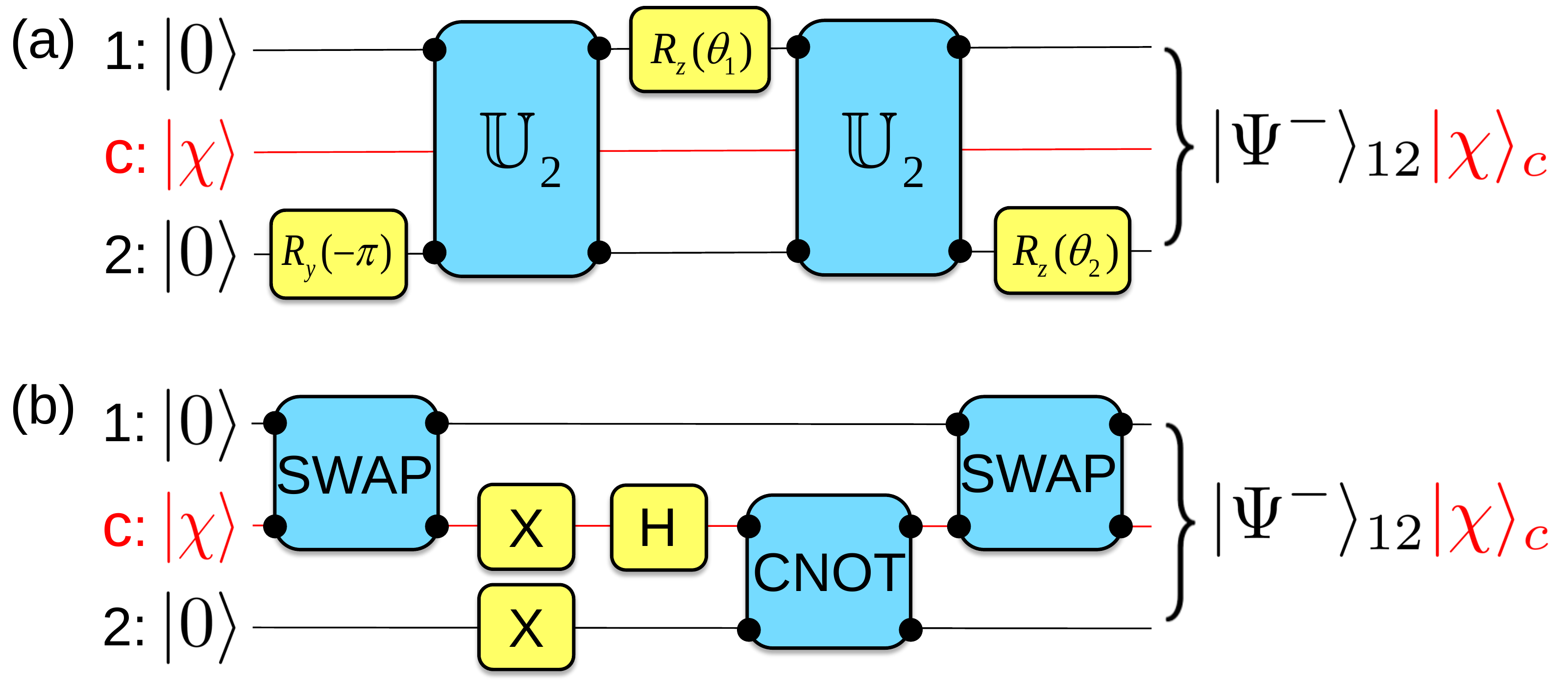}
\caption{(Color online) \label{fig:Bell} 
(a) A quantum circuit for generating the singlet Bell state $\vert\Psi^-\rangle=\frac{1}{\sqrt{2}}(\vert 01\rangle-\vert 10\rangle)$ (up to a global phase), using the mediated gate $\mathbb{U}_2$, given in Eq.~(\ref{eq:U2def}).
Here, the single qubit rotations are defined as $R_\alpha(\theta)=e^{-\frac{i}{2}\theta\sigma_\alpha}$, where $\alpha=x,y,z$.
The rotation angles are given by $\theta_1=-\arccos(\frac{1}{3})$, $\theta_2=-\frac{1}{6}\pi - \arctan\frac{4\sqrt{2}-3\sqrt{3}}{5}$. 
(b) An efficient Bell-state protocol between nonproximal qubits, based on nearest-neighbor, pairwise gates. 
Here, \textsc{h} is the Hadamard gate and \textsc{x} is the Pauli gate $\sigma_x$, corresponding to a $\pi$ rotation about the $x$ axis of the Bloch sphere (up to a global phase).
In this figure, and several other subsequent figures, we note that the central ancilla spin $c$ mediates the multi-qubit gates.
The initial state of the ancilla $|\chi\rangle$ is arbitrary, and it returns to its initial state at the end of the operation.
For completeness, we include $c$ in these circuit diagrams and use filled circles to indicate the qubits being acted on.}
\end{figure}

We have solved Eq.~(\ref{eq:rep}) numerically, obtaining several two-qubit states of interest. 
Our procedure involves maximizing the state fidelity,
$f= |\langle \psi_\text{des}|\psi_\text{actual} \rangle|^2 $, where $|\psi_\text{actual} \rangle$ is the outcome of Eq.~(\ref{eq:rep}), and $|\psi_\text{des} \rangle$ is the desired outcome.  
The two-qubit mediated gate used in the simulations is given by Eq.~(\ref{eq:U2def}), and the individual single-qubit rotations $U_1$ are determined using global optimization methods, as described in Appendix~B.
In principle, we could also allow the circuit depth $n$ to vary.
However, we find that maximally entangled Bell states can already be obtained when $n=2$.  
The resulting circuit for the singlet Bell state $\vert\Psi^-\rangle=\frac{1}{\sqrt{2}}(\vert 01\rangle-\vert 10\rangle)$ is shown in Fig.~\ref{fig:Bell}(a). Other Bell states can be generated in a similar fashion. 

We can compare our mediated gate protocol to the conventional Bell-state protocol based on nearest-neighbor gates, as shown in Fig.~\ref{fig:Bell}(b). 
In the latter case, \textsc{cnot} is used to generate the Bell state, while the \textsc{swap} gates are used to make the qubits proximal.
The minimal circuit depth needed to construct \textsc{cnot} is 2~\cite{Loss98}. 
Comparing Figs.~\ref{fig:Bell}(a) and \ref{fig:Bell}(b), we obtain an exchange gate circuit depth of $n=2$ for the mediated gate protocol and $n=4$ for the \textsc{swap}-based protocol.
The mediated gate therefore offers distinct advantages for generating arbitrary states.

\begin{figure*}[t]
\centering
\includegraphics[width=5.5in]{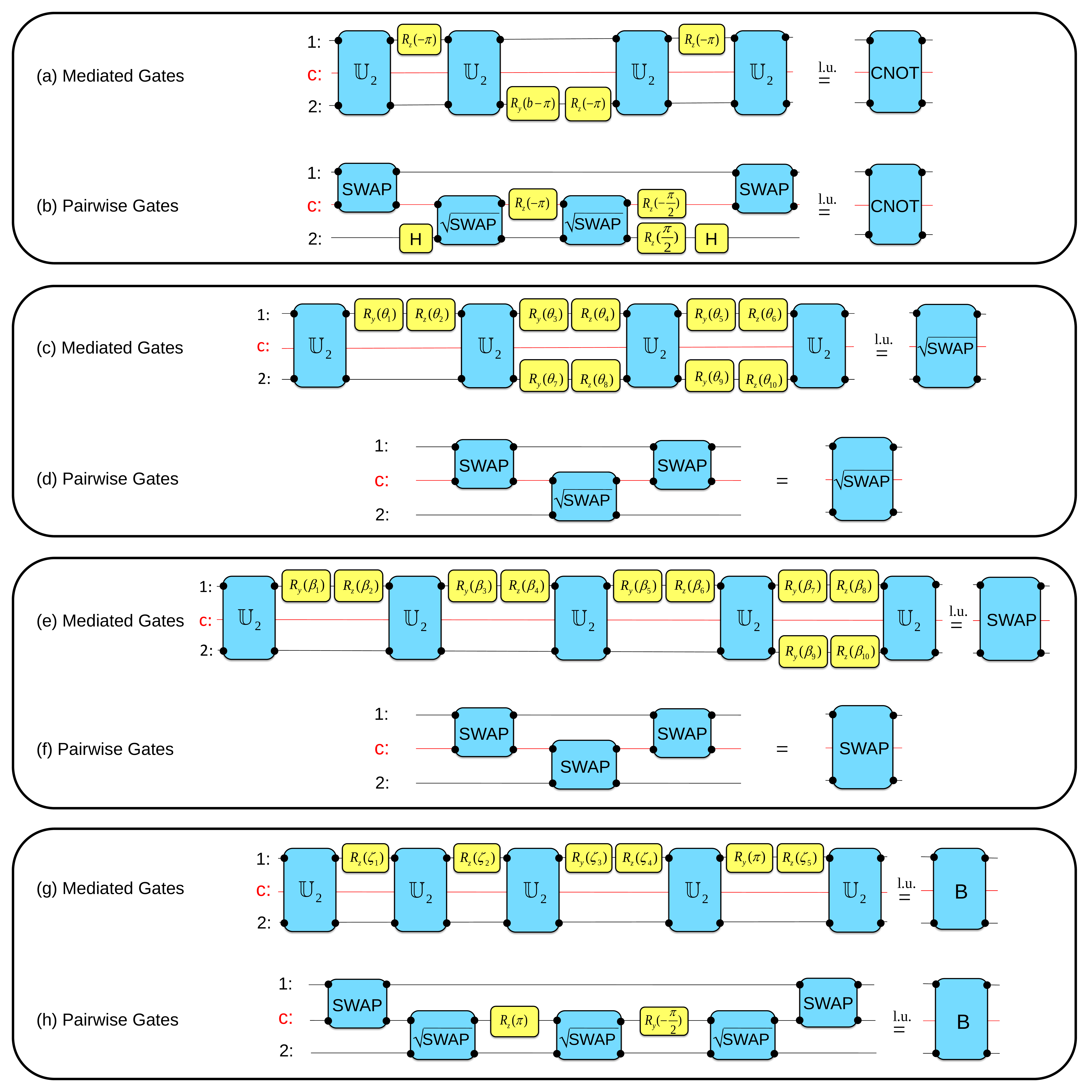}
\caption{(Color online) \label{fig:Ubus2gates} 
A comparison of circuits used to construct some common quantum gates.  
Each case presents two results: a mediated gate construction, obtained using global optimization methods (as described in Appendix~B), and a ``conventional" circuit, based on nearest-neighbor, pairwise gates.
The pairwise gates require extra \textsc{swap} operations when the qubits being acted on are not proximal.
In the cases shown here, the qubits (1 and 2) are separated by one ancilla ($c$).
(a), (b) \textsc{cnot} gates.  
For the mediated gate circuit, we have $b=-\arccos(-1/3)$.
The pairwise gate circuit is given in~\cite{Loss98}.
(c), (d) $\sqrt{\textsc{swap}}$ gates. 
For the mediated gate circuit, we have $\theta_1 = 0.524\pi$, $\theta_2 = 0.549\pi$, $\theta_3 = 1.015\pi$, $\theta_4 = 0.100\pi$, $\theta_5 = 0.392\pi$, $\theta_6 = -0.305\pi$, $\theta_7 = -0.437\pi$, $\theta_8 = 0.626\pi$, $\theta_9 = -0.906\pi$, and $\theta_{10} = -0.174\pi$. 
(These parameters can be obtained up to machine precision.  Here, and elsewhere throughout the paper, we halt the optimization procedure when the objective function is smaller than $10^{-14}$.)
For the pairwise gate circuit, we note that  $\textsc{swap}$ and  $\sqrt{\textsc{swap}}$ are ``natural" gates for spin qubits, whose interactions are of the isotropic Heisenberg type.
As a result, the pairwise gate circuits for $\textsc{swap}$ and  $\sqrt{\textsc{swap}}$ are very simple.
(e), (f) \textsc{swap} gates. 
For the mediated gate circuit, we have $\beta_1 = -0.737\pi$, $\beta_2 = -0.465\pi$, $\beta_3 = -0.543\pi$, $\beta_4 = 0.700\pi$, $\beta_5 = 0.807\pi$, $\beta_6 = 0.009\pi$, $\beta_7 = -0.278\pi$, $\beta_8 = 0.369\pi$, $\beta_9 = 0.274\pi$, and $\beta_{10} = -0.325\pi$.
(g), (h) \textsc{b} gates. 
For the mediated gate circuit, we have $\zeta_1=0.297\pi$, $\zeta_2=0.788\pi$, $\zeta_3=0.660\pi$, $\zeta_4=-1.092\pi$, and $\zeta_5=0.579\pi$. 
For the pairwise gate circuit, the circuit was constructed by first solving for the \textsc{b} gate in terms of $\sqrt{\textsc{swap}}$ gates, using the global optimization methods described in Appendix~B.
\textsc{swap} gates were then applied, to make the qubit states proximal. }
\end{figure*}

\subsection{Experimental proposal for a triple quantum dot}
\label{sec:Experiment}

Triple quantum dots have been investigated in several laboratories~\cite{Laird10,Gaudreau11}.
Here, we suggest a specific protocol for generating a Bell state in a triple quantum dot, using the mediated gate protocol in Fig.~\ref{fig:Bell}(a). 
Our proposal includes the supporting initialization and verification steps, and it is based on existing experimental methods.
We note that Bell states can also be produced via standard, conventional (\emph{i.e.}, nonmediated) techniques~\cite{Petta05}.
The purpose of this section is simply to outline a proof-of-principle experiment that employs mediated gates.

To generate a Bell state using  mediated gates, we must first initialize the triple dot into the separable state $\ket{0}_1\ket{\chi}_c\ket{0}_2$, as shown on the left in Fig.~\ref{fig:Bell}(a). 
There are two common procedures for initializing quantum dot spin qubits:  the preferential loading of single-electron spin ground states ($|0\rangle_1$ and $|0\rangle_2$) in a large magnetic field~\cite{ElzermanAPL04}, and the preferential loading of a two-electron singlet state ($|S\rangle_1$)~\cite{Petta05}. 
The latter can be transformed into the spin ground state of a double quantum dot ($|0\rangle_1|0\rangle_c$) by adiabatically detuning the double dot in a moderate magnetic field~\cite{Petta05}.
Both of these methods require a magnetic field, and we have confirmed that the protocol shown in Fig.~\ref{fig:Bell}(a) is unaffected by a uniform field, up to an overall phase factor.
For the singlet loading method, the desired initial state is achieved, finally, by performing a \textsc{swap} operation between qubit $c$ and qubit 2.
Once the triple dot has been initialized, the mediated gate protocol is implemented as shown in Fig.~\ref{fig:Bell}(a), giving the result $|\Psi^-\rangle_{12}|\chi\rangle_c$. 

The verification step is performed most conveniently via spin-to-charge conversion, using a singlet projection procedure~\cite{Petta05}.
We first perform a \textsc{swap} operation between qubit $c$ and qubit 2, so the two spins in the singlet state become proximal: 
$|\Psi^-\rangle_{12}|\chi\rangle_c\rightarrow|\Psi^-\rangle_{1c}|\chi\rangle_2$.
Dots 1 and $c$ are then detuned, so that the electron in dot $c$ tunnels to dot 1 only if the two electrons form a singlet, due to the large singlet-triplet energy splitting in a single quantum dot.
This projection technique requires a moderate (not too large) magnetic field, so that the singlet remains the ground state of the two-electron dot.

\subsection{Construction of arbitrary gates}
We now consider protocols for generating arbitrary two-qubit gates, using the mediated gate $\mathbb{U}_2$ as an entangling resource, in combination with arbitrary single-qubit gates $U_1$. 
We adopt a strategy analogous to Eq.~(\ref{eq:rep}), given by
\begin{equation}
U_2 = (U_1\otimes U_1)[\mathbb{U}_2  (U_1\otimes U_1)]^n. \label{eq:gates}
\end{equation}
As before, we solve this equation using global optimization techniques, as described in Appendix~B.
The results for some familiar gates are shown in Figs.~\ref{fig:Ubus2gates}(a), \ref{fig:Ubus2gates}(c), \ref{fig:Ubus2gates}(e), and \ref{fig:Ubus2gates}(g).
These results appear to have the smallest possible circuit depth, based on exhaustive searches.
None of the gates requires more than five applications of $\Ub_2$.

Our result for \textsc{cnot} is indicated in Fig.~\ref{fig:Ubus2gates}(a).
This mediated gate circuit employs four $\Ub_2$ gates.
The corresponding circuit for conventional, pairwise gate operations employs two $\sqrt{\textsc{swap}}$ gates when the qubits are proximal~\cite{Loss98}.
When the qubits are nonproximal, additional \textsc{swap} gates are needed, as indicated in Fig.~\ref{fig:Ubus2gates}(b).
Thus, for the second-nearest-neighbor geometry shown in Fig.~\ref{fig:1+2}, the mediated and conventional \textsc{cnot} circuits have equal circuit depths, with $n=4$.

Figure~\ref{fig:Ubus2gates} also shows mediated gate results for several other types of gates, as well as the corresponding conventional, pairwise gate circuits.
For the examples shown here, the mediated gate method has equal or larger circuit depths compared to the pairwise gate method.
The examples where the pairwise gate method is more efficient fall into the SWAP family, which is the ``natural" gate for spin qubits, since it is generated by the isotropic Heisenberg interaction.
There are other, less familiar gates for which the mediated gate circuit is more efficient; the gate $\Ub_2$ is an obvious example.
Generally, we expect that the mediated gate method should be more likely to improve the circuit depth of larger gates (\emph{e.g.}, Toffoli) when multiqubit entangling gates like $\Ub_3$ are available, or when the central spin can be replaced with a spin bus.
We discuss both of these examples below.

There are several well-known techniques for constructing arbitrary two-qubit gates, which can be adapted for mediated gates.  
The most efficient method involves the so-called \textsc{b} gate~\cite{JZhang04prl}, which is defined as the only two-qubit gate that can generate generic two-qubit operations from two successive applications.
Figure~\ref{fig:Ubus2gates}(g) shows our globally optimized circuit for a \textsc{b} gate, which employs five $\Ub_2$ gates.
This result (together with \cite{JZhang04prl}), constitutes a formal proof that $n\leq 10$ for a mediated gate with optimal circuit depth, as described in Eq.~(\ref{eq:gates}).
It also forms a constructive protocol for generating an arbitrary two-qubit gate using 10 $\Ub_2$ gates.
However, we note that the bound $n\leq 10$ does not appear to be tight, since none of the gates we have solved requires more than five applications of $\Ub_2$.

To conclude this section, we consider the scaling properties of the two-qubit mediated gate scheme for a spin bus geometry~\cite{Friesen07}. 
Specifically, we consider an odd-size spin chain of length $N$, and two external qubits. 
When the bus is constrained to its ground-state energy manifold, it can be treated as a spin-1/2 pseudospin~\cite{Friesen07}.
The effective interaction between the qubits and the bus pseudospin has a Heisenberg form~\cite{Shim11, Oh11}, with an effective coupling constant $J^*\propto J/\sqrt{N}$~\cite{Friesen07}. 
We can immediately apply all our three-qubit protocols, simply by replacing the central qubit in Fig.~\ref{fig:1+2} with a bus and replacing $J$ with $J^*$ when we calculate the gate period $T_g$. 
The resulting bus gate $\Ub_2$ is identical to the two-qubit mediated gate, and the protocols proceed as before, except that the qubits can now be far apart.
The exchange gate circuit depth for the bus protocol is the same as that for mediated gates.  
Specifically, it is independent of $N$.
The gate period $T_g$ scales as $\sqrt{N}$, however, since $T_g\propto 1/J^*$.
In contrast, the circuit depth of a conventional gate protocol, based on pairwise \textsc{swap} gates, is proportional to $N$, while $T_g$ is independent of $N$ for a given pairwise operation.
Thus, the spin bus architecture has much better scaling properties than the conventional gate protocol, in terms of both total gate time [$O(\sqrt{N})$ \emph{vs}.\ $O(N)$] and circuit depth [$O(1)$ \emph{vs}.\ $O(N)$], with immediate consequences for quantum error correction~\cite{Svore05}.

\section{Three-Qubit Mediated Gate, $\Ub_3$}
\subsection{Mediated gate, $\mathbb{U}_3$}
We now consider the mediated gate geometry shown in Fig.~\ref{fig:Ubus3}(a), with three qubits coupled through a single mediating spin, $c$.
The system Hamiltonian is given by
\begin{equation}
H=J_{1c} \mathbf{s}_1\cdot \mathbf{s}_c+J_{2c}\mathbf{s}_2\cdot \mathbf{s}_c
+J_{3c}\mathbf{s}_3\cdot \mathbf{s}_c ,
\label{eq:H3+1}
\end{equation}
and the time evolution operator is given by $U(t)=e^{-iHt}$.
If qubits 1--3 are arranged in a linear geometry rather than the ``star" geometry shown in Fig.~\ref{fig:Ubus3}, then an effective star geometry can still be achieved by introducing a spin bus architecture~\cite{Friesen07}, where $c$ is an odd-size bus.

For larger geometries, the group theoretical methods described in Appendix~A become cumbersome.
However, for the special case of equal couplings, $J=J_{1c}=J_{2c}=J_{3c}$, we can still obtain mediated gates analytically.
We do this by computing $U(t)$ in the angular momentum basis, where it is diagonal~\cite{Friesen07}. 
We then transform it to the computational basis and identify the gate periods $t=T_g$ for which the special decomposition $U=\Ub_3 \otimes I$ is satisfied.  
Here, $\Ub_3$ is the mediated gate acting on qubits 1--3, while $I$ is the single-qubit identity operator acting on spin $c$.
This procedure produces four different mediated gates~\cite{Friesen07}.
The first gate is the trivial identity operator, obtained at the gate periods $T_g=(8m) \pi/J$ ($m$ is an integer).
The second gate occurs at the gate periods $T_g = (8m+2)\pi/J$, and takes the form
\begin{widetext}
\begin{equation}
\Ub_3 =
i\left( {\begin{array}{cccccccc}
 1 & 0 & 0 & 0 & 0 & 0 & 0 & 0  \\
 0 & -1/3 & 2/3 & 0 & 2/3 & 0 & 0 & 0  \\
 0 & 2/3 & -1/3 & 0 & 2/3 & 0 & 0 & 0  \\
 0 & 0 & 0 & -1/3 & 0 & 2/3 & 2/3 & 0  \\
 0 & 2/3 & 2/3 & 0 & -1/3 & 0 & 0 & 0  \\
 0 & 0 & 0 & 2/3 & 0 & -1/3 & 2/3 & 0  \\
 0 & 0 & 0 & 2/3 & 0 & 2/3 & -1/3 & 0  \\
 0 & 0 & 0 & 0 & 0 & 0 & 0 & 1  \\
 \end{array} } \right).
\end{equation}
\end{widetext}
The third gate occurs at the gate periods $T_g = (8m+4)\pi/J$, and is given by $\Ub_3^2=-I$.
The fourth gate occurs at the gate periods $T_g = (8m+6)\pi/J$, and is given by $\Ub_3^3=-\Ub_3$.

\subsection{Generation of arbitrary states}
\begin{figure}[t]
\centering
\includegraphics[width=1.0\linewidth]{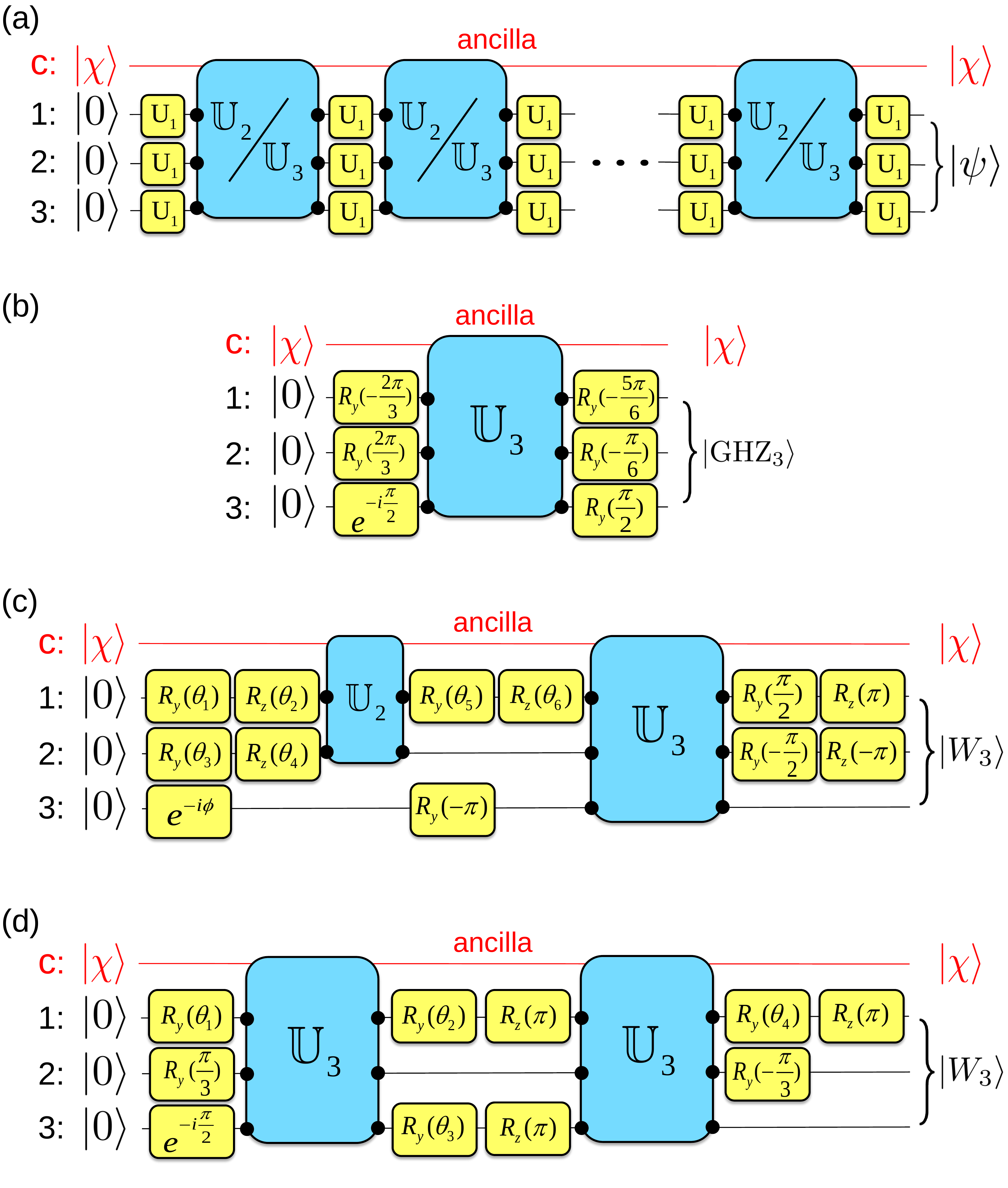}
\caption{(Color online)  (a) A general circuit for generating an arbitrary three-qubit state, using $\Ub_2$ and/or $\Ub_3$ gates. Note that $\Ub_2$ can act on different pairs of qubits.
(b) A circuit for generating a $\ket{\text{GHZ}_3}$ state, using the three-qubit mediated gate $\Ub_3$.
(c) A circuit for generating a $|W_3\rangle$ state, using both $\Ub_2$ and $\Ub_3$ gates, with  $\theta_1=-0.262\pi$, $\theta_2=0.730\pi$, $\theta_3=-1.356\pi$, $\theta_4=0.349\pi$, $\theta_5=1.193\pi$, $\theta_6=0.270\pi$, and $\phi=1.299\pi$. 
(d) An alternative circuit for generating a $|W_3\rangle$ state, using only $\Ub_3$ gates, with $\theta_1=0.529\pi$, $\theta_2=0.725\pi$, $\theta_3=-0.608\pi$, and $\theta_4=-0.137\pi$. }
\label{fig:GHZ3Ubus3}
\end{figure}

The methods used to generate two-qubit states and gates can also be extended to three-qubit problems.
However, three-qubit protocols are slightly more complicated because they can involve two-qubit gates, three-qubit gates, or both.  
The most general scheme for generating a three-qubit state is shown in Fig.~\ref{fig:GHZ3Ubus3}(a).
We note that higher order gates such as the three-qubit mediated gate $\Ub_3$ can potentially achieve shorter circuit depths, because they are more parallel than two-qubit gates.
The global optimization techniques used to solve Eq.~(\ref{eq:rep}) can also be applied to Fig.~\ref{fig:GHZ3Ubus3}(a).  

\begin{figure}[t]
\centering
\includegraphics[width = 3.5in]{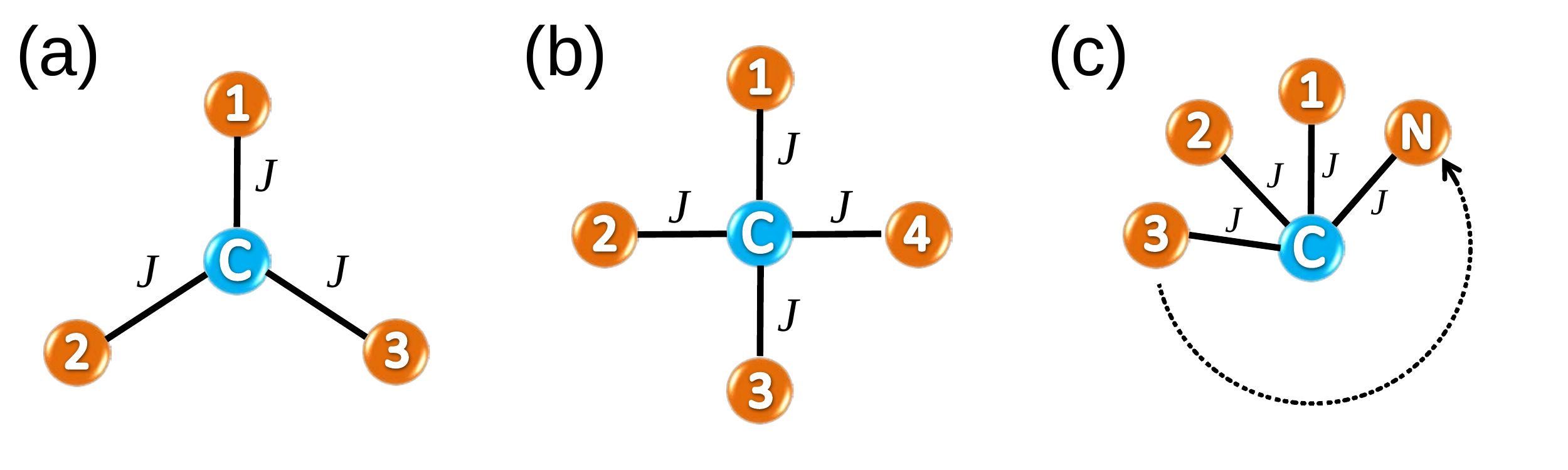}
\caption{(Color online) Multiqubit ``star" geometries for implementing mediated gates.  
Here, the qubits are labeled with numbers and the ancilla spins labeled $c$ mediate the gates.
(a) The $\Ub_3$ gate acts on qubits 1--3 when the three qubit couplings $J$ are equal.
(b) To generate the four-qubit cluster state $|C_4\rangle$, we implement three-qubit mediated gates $\Ub_3$ by turning on the couplings to three of the qubits at a time.
(c) To generate a $W_N$ state, we consider $N$ qubits connected simultaneously to the ancilla spin $c$, with equal couplings $J$.
In each of these geometries, the ancilla spin can be replaced with an odd-size spin bus.
In this case, $c$ represents the pseudospin of the bus ground state~\cite{Friesen07}.}
\label{fig:Ubus3}
\end{figure} 

There are known to be two nonfungible forms of entanglement for three qubits~\cite{Dur00}:
the $W$-state family, characterized by the symmetric form 
\begin{equation}
|W_3\rangle =\frac{1}{\sqrt{3}} (|001\rangle+|010\rangle+|100\rangle ) ,
\end{equation} 
and the 
Greenberger-Horne-Zeilinger (GHZ)-state family~\cite{GHZ}, characterized by the symmetric form
\begin{equation}
|\text{GHZ}_3\rangle = \frac{1}{\sqrt{2}} (|000\rangle +|111\rangle) .
\end{equation} 
The GHZ state is understood to be maximally entangled for three qubits.

We have applied global optimization methods to obtain $|\text{GHZ}_3\rangle$ and $|W_3\rangle$, obtaining the results shown in Figs.~\ref{fig:GHZ3Ubus3}(b)-\ref{fig:GHZ3Ubus3}(d). 
For $|W_3\rangle$, we provide two different strategies. 
One uses a combination of $\Ub_2$ and $\Ub_3$; the other uses $\Ub_3$ only.
They both have the same circuit depth, $n=2$.
Remarkably, we find that the GHZ state can be attained using $\Ub_3$ as the only entangling resource with just a single application:
\begin{equation}
\ket{\text{GHZ}_3}= (U_1\otimes U_1\otimes U_1) \Ub_3 (U_1\otimes U_1\otimes U_1)\ket{000} .
\end{equation} 
The circuit is optimal ($n=1$), indicating that $\Ub_3$ is a perfect entangler for the three-qubit GHZ state family~\cite{Dur00}.
We can compare this result to the conventional, pairwise gating circuit for $|\text{GHZ}_3\rangle$, which uses two \textsc{cnot} gates~\cite{Mermin}.
In a quantum dot quantum computer, this would require at least four exchange gate operations, or $n=4$.
It is interesting to note that $\Ub_3$ is locally equivalent to the time evolution operator describing the three-qubit triangular geometry (evaluated at a special time)~\cite{Galiautdinov08}.
The latter gate is also capable of generating $|\text{GHZ}_3\rangle$ in a single time step.

Although $\Ub_3$ acts on just three qubits at a time, it is interesting to note that it can also be used as an entangling resource for larger systems. 
For example, we can consider cluster states, which represent an important entanglement family used for one-way quantum computing~\cite{Briegel01, Raussendorf01}. 
The four-qubit cluster state $\vert C_4\rangle$ is defined as
\begin{equation}
\vert C_4\rangle \eq \frac{1}{2}(\vert 0000\rangle + \vert 0011\rangle + \vert 1100\rangle - \vert 1111\rangle) .
\end{equation}

We have solved $\vert C_4\rangle$ numerically, for the geometry shown in Fig.~\ref{fig:Ubus3}(b).
Here, the ancilla spin $c$ can be connected to each of the four qubits.
However, we assume that only three of the couplings are turned on at a time.
For example, $\Ub_3(1,2,3)$ indicates that the couplings between $c$ and qubits 1--3 are turned on, thus implementing the gate $\Ub_3$ between those three qubits.
Hence, we obtain a numerical solution for $\vert C_4\rangle$ of the form
\begin{align}
\vert C_4\rangle= & U_1^{\otimes 4}\Ub_3(1,2,3) U_1^{\otimes 4}\Ub_3(1,2,4) U_1^{\otimes 4}\Ub_3(1,2,3) \notag\\
                  &\times U_1^{\otimes 4}\Ub_3(2,3,4) U_1^{\otimes 4} \ket{0000} .
\end{align}
Here, $U_1^{\otimes 4}$ represents arbitrary single-qubit rotations acting on each of the four qubits.
According to our definition of circuit depth, this protocol corresponds to $n = 4$. 

We can compare our mediated gate solution to a conventional sequence for generating $|C_4\rangle$, based on nearest-neighbor pairwise gates.
The conventional scheme involves three sequential applications of the phase gate $\text{diag}(1, 1, 1, -1)$, in addition to single-qubit rotations~\cite{Briegel01,Raussendorf01}.  
Since the phase gate is locally equivalent to \textsc{cnot}, it can be decomposed into two exchange gates plus single-qubit rotations.
The resulting circuit depth for the conventional protocol is therefore $n=6$.
Thus, again, we find that mediated gates offer a considerable improvement in terms of circuit depth. 

\subsection{Construction of arbitrary gates}

We now turn to the construction of three-qubit quantum gates using $\Ub_3$.
As an example, we determine an explicit gate sequence for generating the Toffoli gate, defined as 
\begin{equation}
U_T =
\left( {\begin{array}{cccccccc}
 1 & 0 & 0 & 0 & 0 & 0 & 0 & 0  \\
 0 & 1 & 0 & 0 & 0 & 0 & 0 & 0  \\
 0 & 0 & 1 & 0 & 0 & 0 & 0 & 0  \\
 0 & 0 & 0 & 1 & 0 & 0 & 0 & 0  \\
 0 & 0 & 0 & 0 & 1 & 0 & 0 & 0  \\
 0 & 0 & 0 & 0 & 0 & 1 & 0 & 0  \\
 0 & 0 & 0 & 0 & 0 & 0 & 0 & 1  \\
 0 & 0 & 0 & 0 & 0 & 0 & 1 & 0  \\
 \end{array} } \right).
\end{equation}
Our strategy is analogous to the state-generating circuit in Fig.~\ref{fig:GHZ3Ubus3}(a), where we interspersed $\Ub_2$ or $\Ub_3$ gates with arbitrary single-qubit rotations.
Our best result for constructing the Toffoli gate by this method is a gate sequence containing five $\Ub_2$ gates and seven $\Ub_3$ gates, giving a total exchange-gate circuit depth of $n = 12$. 

We can compare our mediated gate solution to a conventional Toffoli gate construction.
A Toffoli circuit using \textsc{cnot} gates as the entangling resource has been presented in \cite{Nielsen&Chuang} and \cite{Shende09}; it consists of six sequential \textsc{cnot} gates.
We can decompose this into a sequence of nearest-neighbor exchange gates, including intermediate \textsc{swap} gates when necessary.
After identifying the exchange gates that can be performed in parallel, this procedure gives a circuit depth of $n=16$.
Alternatively, if we allow other two-qubit gates in this procedure, in addition to \textsc{cnot}, it can be shown that five sequential two-qubit gates are necessary and sufficient for implementing a Toffoli gate~\cite{DiVincenzo94}. 
However, some of these gates are decomposed into exchange gate sequences with $n>2$.
Based on such considerations, it appears that the mediated gate circuit with a circuit depth of $n = 12$ for constructing a Toffoli gate is always more efficient than a conventional gate circuit. 

\section{Mediated Gates, $\Ub_{2N+1}$ ($N > 1$)}
\label{sec:Un}
The previous approach to state generation and gate construction using mediated gates can be extended to systems with more than three qubits. 
There are many qubit architectures of interest.
Here, we consider the ``star" geometry shown in Fig.~\ref{fig:Ubus3}(c).
In cases where it is experimentally challenging to fabricate a star geometry, due to physical constraints, it may be convenient to replace the central spin $c$ with an odd-size spin bus~\cite{Friesen07}.
In this case, nontrivial mediated gates can be obtained when an odd number of qubits is simultaneously coupled to the bus.
These multiqubit mediated gates, $\Ub_{2N+1}$, are highly parallel and potentially very efficient. 

\begin{figure}[t]
\centering
\includegraphics[width=1.0\linewidth]{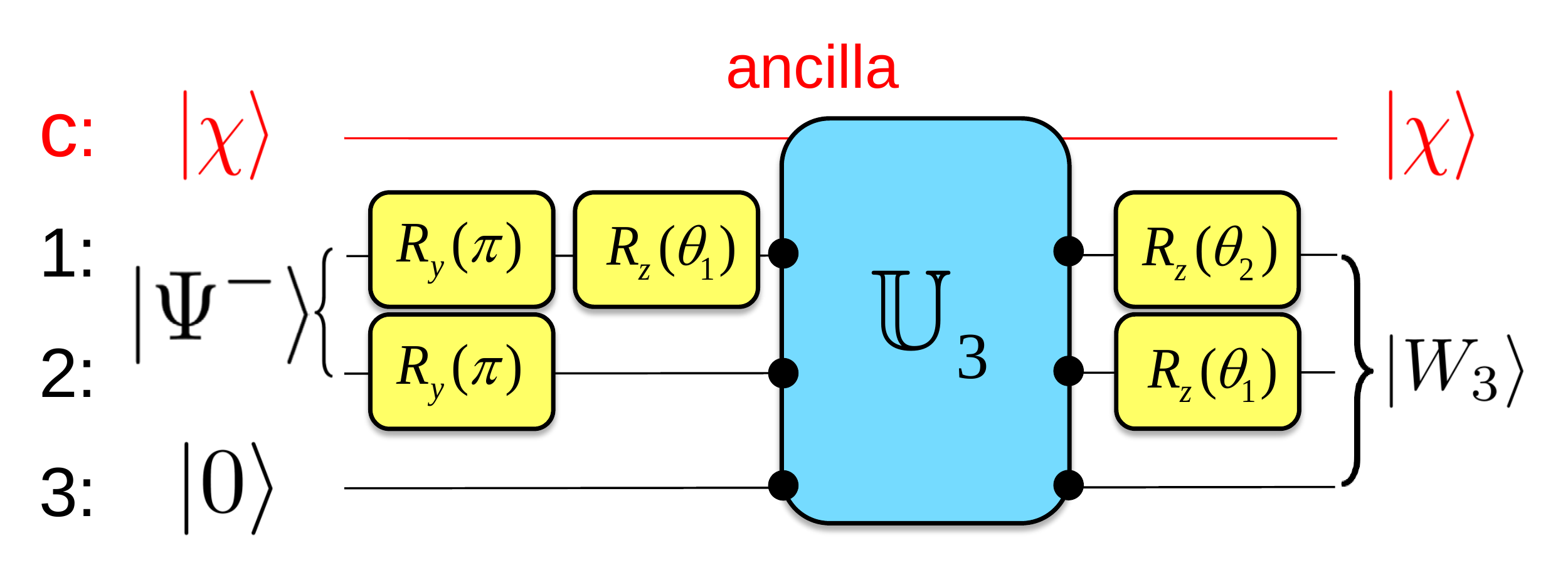}
\caption{(Color online) A mediated gate circuit for generating $\vert W_3\rangle$ using a Bell state as input. 
Here, $\theta_1 = -\theta_2 = \arccos(1/4)$. }
\label{fig:W3_Bell}
\end{figure}

Here, we demonstrate that multiqubit $W$ states can be generated using mediated gates, with very small circuit depths.
The $N$-qubit $W$ state is defined as
\begin{equation}
\vert W_N\rangle = \frac{1}{\sqrt{N}}(\vert 00\ldots 01\rangle + \vert 00\ldots 10\rangle + \ldots + \vert 10\ldots 00\rangle) .
\end{equation}
In Figs.~\ref{fig:GHZ3Ubus3}(c) and \ref{fig:GHZ3Ubus3}(d) we indicate two methods for generating $|W_3\rangle$.
An alternative method is shown in Fig.~\ref{fig:W3_Bell}.
This circuit requires a maximally entangled Bell state, $|\Psi^-\rangle$, as input.
The total circuit depth for this solution ($n=3$) is larger than in Figs.~\ref{fig:GHZ3Ubus3}(c) and \ref{fig:GHZ3Ubus3}(d) because the circuit depth for generating $|\Psi^-\rangle$ is $n=2$.
However, the scheme has the advantage that it may be scalable for odd-size $W$ states.

For the cases $N=1$--3, we have numerically verified the result that
\begin{equation}
\vert W_{2N+1}\rangle = (U_1\otimes U_1) \Ub_{2N+1} (U_1\otimes U_1)\vert \Psi^-\rangle\vert 0\rangle^{\otimes (2N-1)} ,
\end{equation}
which includes the result in Fig.~\ref{fig:W3_Bell} for the case $N=1$.
For all cases, we note that the single-qubit rotations are applied only to qubits 1 and 2 (the qubits in the Bell state).
For the cases of $N = 2, 3$, the generating circuits are similar to Fig.~\ref{fig:W3_Bell}, but with different angles $\theta_1 $ and $\theta_2$.
For each of these cases, the circuit depth is given by $n=3$.

A related, probabilistic scheme can be used to generate the even-size $W$ states.
We first generate the odd-size $W$ state, as described above.
This state can be expressed as
\begin{equation}
\vert W_{2N+1}\rangle = \frac{1}{\sqrt{2N+1}}\vert 0\rangle^{\otimes 2N} \vert 1\rangle + \sqrt{\frac{2N}{2N+1}}\vert W_{2N}\rangle\vert 0\rangle .
\end{equation}
Hence, if one of the qubits is measured in the $z$ basis, with outcome 0, then the state of the remaining qubits will collapse to $|W_{2N}\rangle$.
When $N$ is large, this protocol is successful with a high probability, $P = 2N/(2N+1)$.

\section{Summary and Conclusions}

\begin{table}[h]
\caption{Comparison of the exchange gate circuit depths ($n$) for generating some common quantum states and gates, using two different gating protocols.
Schemes we consider here are based on (i) mediated gates (as described in this paper), or (ii) conventional pairwise gates.  
The pairwise gating method requires extra \textsc{swap} gates when the qubits being acted on are not proximal.
(A useful reference for the scaling of $\vert W_N\rangle$ state using pairwise gates is~\cite{Bodoky07}.)}
\label{table:GateCompare}
\begin{tabular}{lcc}
\hline\hline
State or gate   & \hspace{.2in}No. mediated gates  \hspace{.2in}   & No. pairwise gates\\
\hline
$\vert\Psi^-\rangle$          & 2  & 4 \\
$\vert W_3\rangle$            & 2  & 2 \\
$\vert W_N\rangle$            & 3  & $N-1$ \\
$\vert \text{GHZ}_3\rangle$   & 1  & 4 \\
$\vert C_4\rangle$            & 4  & 6  \\
$\textsc{cnot}$               & 4  & 4  \\
$\sqrt{\textsc{swap}}$        & 4  & 3  \\
$\textsc{swap}$               & 5  & 3  \\
$\textsc{b}$                  & 5  & 5 \\
Toffoli                       & 12 & 16  \\
\hline\hline
\end{tabular}
\end{table}

In this paper, we developed the concept of a \emph{mediated gate} between nonproximal qubits.
This gate is implemented by coupling the qubits simultaneously through a central, ancilla qubit, which is restored to its initial state at the end of the operation. 
We have focused on two and three-qubit gates, although higher dimensional gates can be obtained in similar fashion.
We investigated protocols, based on global optimization techniques, for generating arbitrary states and gates, using mediated gates as the sole entangling resource.  

Several promising results were obtained using mediated gates, as summarized in Table~I.
We showed that a maximally entangled Bell state can be achieved with just two applications of a mediated gate $\Ub_2$, and we proposed an experimental protocol for implementing this procedure in a triple quantum dot.
We showed that several important two-qubit quantum gates can be obtained using five or fewer mediated gates, and we proved that ten exchange gates is the maximum needed for generating an arbitrary two-qubit gate.
We showed how the central ancilla qubit can be replaced with a spin bus, leading to significant improvements in scaling properties, for both the total gate time and the circuit depth.
We also considered the mediated gates $\Ub_N$ with $N \geq 3$, and showed how mediated gate methods might be generalized to higher dimensions.

We find that mediated gates compare favorably with conventional, pairwise gating schemes, which make use of SWAP gates when qubits are not proximal.
For each of the results reported in Table~I, we compare the circuit depths based on mediated gates to those involving conventional pairwise gates.

We thank Jun Zhang and Andrei Galiautdinov for helpful discussions. This work was supported by the DARPA QuEST program through a grant from AFOSR, by NSA/LPS through grants from ARO (W911NF-09-1-0393, W911NF-08-1-0482, and W911NF-12-1-0607), by NSF-PIF (PHY-1104672), and by the United States Department of Defense. The US government requires publication of the following disclaimer: the views and conclusions contained in this document are those of the authors and should not be interpreted as representing the official policies, either expressly or implied, of the US Government.  

\renewcommand{\theequation}{\Alph{section}\arabic{equation}}

\begin{appendix}

\setcounter{section}{0}
\setcounter{equation}{0}

\section{EXISTENCE PROOF FOR $\Ub_2$}
\label{sec:U2Proof}
\begin{figure}[h]
  \includegraphics[width=0.5\linewidth]{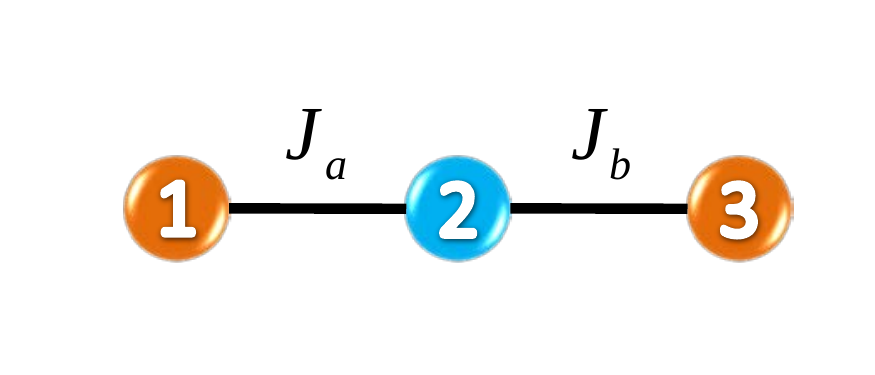}
  \caption{(Color online) Two-qubit mediated gate geometry. Here, ancilla qubit 2 mediates gate $\Ub_2$, which acts on qubits 1 and 3.}
  \label{fig:1+2Ubus}
\end{figure}

Here, we prove that the gate $\Ub_2$, presented in Eq.~(\ref{eq:U2def}) of the main text (and its family), represents the only solutions to the mediated gate problem for two qubits.
For convenience, we adopt slightly different notation than in the main text, as indicated in Fig.~\ref{fig:1+2Ubus}.  
Spins 1 and 3 are the two nonproximal qubits, while spin 2 is the central ancilla qubit. 

We consider the following Hamiltonian for a linear three-qubit array: 
\begin{equation}
H=J_a\, \mathbf{s}_1\cdot \mathbf{s}_2+J_b\, \mathbf{s}_2\cdot \mathbf{s}_3,
\label{eq:Hsupp}
\end{equation}
where $\mathbf{s}_j$ is the spin operator for qubit $j$.
In principle, $J_a$ and $J_b$ may take any value.  
However, we limit our search to the case where the couplings are turned on and off simultaneously.
$J_a$ and $J_b$ are therefore constant throughout the gate operation.
The goal of this Appendix is to identify specific relations between $J_a$ and $J_b$ that lead to mediated gates. 

We make use of the identity~\cite{Dirac, Feynman}
\begin{equation}
4\,\mathbf{s}_i \cdot \mathbf{s}_j=2p^{ij}-I,
\end{equation}
where $p^{ij}$ is the \textsc{swap} (i.e., transposition) operator between spin $i$ and spin $j$, and $I$ is the two-qubit identity operator.
Hamiltonian (\ref{eq:Hsupp}) can then be rewritten as
\begin{equation}
H=\frac{1}{2}(J_a p^{12}+J_bp^{23})-\frac{1}{4}(J_a+J_b).
\end{equation}

The time evolution operator is given by
\begin{align}
U(t) &= e^{-iHt} \nonumber \\
     &=e^{i(J_a+J_b)t/4}e^{-iQJ_bt/2} \nonumber \\
     &=e^{i(J_a+J_b)t/4}\sum_{n=0}^{\infty}\frac{(-iJ_bt/2)^n}{n!}Q^n, \label{eq:Uexpand}
\end{align}
where $\hbar=1$ and we have defined
\begin{equation}
Q\equiv p^{23}+Jp^{12}, 
\end{equation}
with $J= J_a/J_b$.
Since $p^{23}$ and $p^{12}$ are generators of the symmetric group S$_3$,
we may expand $Q^n$ in terms of the S$_3$ group elements:
\begin{equation}
Q^n=a_n p^{231} + b_n p^{312} + c_n p^{12} + d_n p^{13}+ e_n p^{23} + f_n I.
\end{equation} 
Here, $p^{ijk}$ is the tripartite, cyclic permutation operator. 

\begin{table}[h]
\caption{(Color online) Cayley table for the symmetric group S$_3$.}
\label{table:S3Cayley}
\begin{ruledtabular}
\begin{tabular}{ccccccc}
          & $I$       & $p^{12}$ & $p^{13}$ & $p^{23}$ & $p^{231}$ & $p^{312}$\\
\hline
$I$         & $I$         & $p^{12}$ & $p^{13}$ & $p^{23}$ & $p^{231}$ & $p^{312}$ \\
$p^{12}$ & $p^{12}$ & $I$        & $p^{231}$ & $p^{312}$ & $p^{13}$ & $p^{23}$ \\
$p^{13}$ & $p^{13}$ & $p^{312}$ & $I$         & $p^{231}$ & $p^{23}$ & $p^{12}$ \\
$p^{23}$ & $p^{23}$ & $p^{231}$ & $p^{312}$ & $I$         & $p^{12}$ & $p^{13}$ \\
$p^{231}$ & $p^{231}$ & $p^{23}$ & $p^{12}$ & $p^{13}$ & $p^{312}$ & $I$ \\
$p^{312}$ & $p^{312}$ & $p^{13}$ & $p^{23}$ & $p^{12}$ & $I$         & $p^{231}$ \\
\end{tabular}
\end{ruledtabular}
\end{table}

The full set of S$_3$ group operations is listed in Table.~\ref{table:S3Cayley}.
We then deduce the recursion relations for $Q^{n+1}=QQ^{n}$:
\begin{align}
&a_{n+1}=c_n+Jd_n,\\
&b_{n+1}=d_n+Je_n,\\
&c_{n+1}=a_n+Jf_n,\\
&d_{n+1}=b_n+Ja_n,\\
&e_{n+1}=f_n+Jb_n,\\
&f_{n+1}=e_n+Jc_n.
\end{align}
These relations can be expressed compactly as
\begin{equation}
\mathbf{v_{n+1}} = T\mathbf{v_n}, \label{eq:recursion}
\end{equation}
where
\begin{equation}
\mathbf{v_n} = [f_n\ c_n\ e_n\ a_n\ b_n\ d_n]^T ,
\end{equation}
and
\begin{equation}
T = \left(
\begin{array}{cccccc}
0 & J & 1 & 0 & 0 & 0 \\
J & 0 & 0 & 1 & 0 & 0 \\
1 & 0 & 0 & 0 & J & 0 \\
0 & 1 & 0 & 0 & 0 & J \\
0 & 0 & J & 0 & 0 & 1 \\
0 & 0 & 0 & J & 1 & 0
\end{array} \right).
\end{equation}

We now solve the recursion problem analytically.
The $n=0$ term of the summation in Eq.~(\ref{eq:Uexpand}) corresponds to the initial condition $\mathbf{v_0} = [1\ 0\ 0\ 0\ 0\ 0]^T$. 
Equation~(\ref{eq:recursion}) then leads to
\begin{align}
&a_n=\frac{1}{3}[(1+J)^n-(1-J+J^2)^{\frac{n}{2}}],\\
&b_n=\frac{1}{3}[(1+J)^n-(1-J+J^2)^{\frac{n}{2}}],\\
&f_n=\frac{1}{3}[(1+J)^n+2(1-J+J^2)^{\frac{n}{2}}], \\
&c_n=d_n=e_n=0, 
\end{align}
when $n$ is even, and 
\begin{align}
&c_n=\frac{1}{3}[(1+J)^n+(2J-1)(1-J+J^2)^{\frac{n-1}{2}}],\\
&d_n=\frac{1}{3}[(1+J)^n-(1+J)(1-J+J^2)^{\frac{n-1}{2}}],\\
&e_n=\frac{1}{3}[(1+J)^n+(2-J)(1-J+J^2)^{\frac{n-1}{2}}], \\
& a_n=b_n=f_n=0,
\end{align}
when $n$ is odd.
Performing the sum over $n$, the time evolution operator can finally be written as
\small
\begin{widetext}
\begin{align} \label{eq:bigU}
U&(t)=\frac{e^{iJ_bt(1+J)/4}}{3}\notag
\Biggl( p^{231} \left\{ \cos[J_bt(1+J)/2]-\cos(J_bt\sqrt{1-J+J^2}/2) \right\}
+p^{312} \left\{ \cos[J_bt(1+J)/2]-\cos(J_bt\sqrt{1-J+J^2}/2) \right\}  \\\notag
     &\hspace{0.3in} +I \left\{ \cos[J_bt(1+J)/2]+2\cos(J_bt\sqrt{1-J+J^2}/2) \right\}
     -ip^{12} \left\{ \sin[J_bt(1+J)/2]+\frac{2J-1}{\sqrt{1-J+J^2}}\sin(J_bt\sqrt{1-J+J^2}/2) \right\} \\\notag
     &\hspace{0.3in} -ip^{13} \left\{ \sin[J_bt(1+J)/2]-\frac{1+J}{\sqrt{1-J+J^2}}\sin(J_bt\sqrt{1-J+J^2}/2) \right\} \\
     &\hspace{0.3in} -ip^{23} \left\{ \sin[J_bt(1+J)/2]+\frac{2-J}{\sqrt{1-J+J^2}}\sin(J_bt\sqrt{1-J+J^2}/2) \right\} \Biggr) \normalsize.
\end{align} 
\end{widetext} 

The mediated gates we search for can be decomposed as 
\begin{equation}
\label{eq:Udecomp}
U=\Ub_2 \otimes I ,
\end{equation} 
where $\Ub_2$ acts on qubits 1 and 3, while $I$ is the single-qubit identity operator acting on spin 2.
Condition (\ref{eq:Udecomp}) is satisfied when the coefficients of $p^{231}, p^{312}, p^{12}$, and $p^{23}$ in Eq.~(\ref{eq:bigU}) all vanish.
The solution is given by
\begin{equation}
J=\frac{J_a}{J_b}=1, \label{eq:solJ}
\end{equation} 
with
\begin{align}
&\cos(J_bt/2)=\cos(J_bt), \label{eq:cond1}\\
&\sin(J_bt/2)=-\sin(J_bt). \label{eq:cond2}
\end{align}
We then solve Eqs.~(\ref{eq:cond1}) and (\ref{eq:cond2}) to obtain the mediated gate periods, $t=T_g$:
\begin{equation}
J_bT_g=0,\frac{4\pi}{3},\frac{8\pi}{3},4\pi , \ldots . \label{eq:solTg}
\end{equation}

The time evolution operator obtained from Eqs.~(\ref{eq:solJ})--(\ref{eq:cond2}) is given by
\begin{equation}
U(T_g)=e^{iJ_bT_g/2}[I\cos(J_bt)-ip^{13}\sin(J_bT_g)] .
\end{equation}
Equation~(\ref{eq:solTg}) then leads to three distinct types of gate operations.
When $J_bT_g=(4m)\pi$, with $m$ an integer, we obtain the trivial gate, $U(T_g)=I$.
When $J_bT_g=(4m+\frac{4}{3})\pi$, we obtain the nontrivial result
\begin{equation}
U(T_g)=e^{-i\frac{\pi}{3}}(\frac{1}{2}I-i\frac{\sqrt{3}}{2}p^{1,3}). \label{eq:UTg}
\end{equation}
The decomposition of Eq.~(\ref{eq:Udecomp}) leads to the identification of the mediated gate $\Ub_2$, given in Eq.~(\ref{eq:U2def}).
When $J_bT_g=(4m+\frac{8}{3})\pi$, we obtain the complementary gate $U(T_g)=\Ub_2^2 \otimes I$.
Finally, we note that $\Ub_2^3=I$.
Thus, $\Ub_2$, $\Ub_2^2=\Ub_2^{-1}$, and $I$ comprise the full set of two-qubit mediated gates.

\section{GLOBAL OPTIMIZATION TECHNIQUES FOR CONSTRUCTING QUANTUM STATES AND GATES}

In this Appendix, we outline the global optimization methods used to solve Eqs.~(\ref{eq:rep}) and (\ref{eq:gates}), which act on two qubits.  
Identical methods can also be used to generate states and construct gates involving more than two qubits.

Equations~(\ref{eq:rep}) and (\ref{eq:gates}) can be summarized as follows.
An arbitrary two-qubit quantum circuit is formed of units comprised of one entangling gate, $\Ub_2$, sandwiched between single-qubit unitary rotations.
One or more of these units can be combined, sequentially, to form a circuit.
The single-qubit rotations in this protocol are arbitrary.
However, the entangling gate $\Ub_2$ is fixed, with the form shown in Eq.~(\ref{eq:U2def}).

Three scalar parameters are required, to fully specify an arbitrary single-qubit rotation, up to a global phase factor (\emph{e.g.}, the Euler angle construction).
Here, we adopt the ZYZ decomposition~\cite{Nielsen&Chuang}:
\begin{equation}
U_1(\alpha,\beta,\gamma)=
e^{-i\alpha \sigma_z/2} e^{-i\beta \sigma_y/2} e^{-i\gamma \sigma_z/2} .
\end{equation}
In the most general case, the rotations will be applied to both qubits, before and after each exchange gate.
The construction can be further simplified by noting that terms such as $(U_1\otimes U_1)(U_1\otimes U_1)$ are redundant and can be collapsed into the form $U_1\otimes U_1$.
Thus, up to $6(n+1)$ rotation angles are required, to specify an arbitrary gate sequence of circuit depth $n$.

Here, we employ global optimization techniques, to search through this large parameter space.
We have found that multistart clustering algorithms~\cite{DixonBook,HorstBook} are particularly effective for solving this problem.
We first define an appropriate objective function to be minimized.  
For generating arbitrary states, as in Eq.~(\ref{eq:rep}), we use the infidelity ($1-f$) of the desired final state $|\psi_\text{des} \rangle$ as the objective function, where
\begin{eqnarray}
f &=& |\langle \psi_\text{des}|\psi_\text{actual} \rangle|^2 \\ \nonumber
&=& |\langle \psi_\text{des}|(U_1\otimes U_1)[\Ub_2(U_1\otimes U_1)]^n|00 \rangle|^2 .
\end{eqnarray}
For generating arbitrary gates, as in Eq.~(\ref{eq:gates}), we use the operator error norm $\epsilon$ as the objective function, where
\begin{equation}
\epsilon = ||U_{2,\text{des}}-U_{2,\text{actual}}|| .
\end{equation}

The global optimization is performed in two steps.
In the first step, we use the multistart algorithm to identify potential candidate solutions.  
Then, we use these solutions as a first guess in a local Nelder-Mead downhill simplex search~\cite{NelderMead}.
The final outcome generally provides results with very low or very high accuracy.
The latter are accepted as valid solutions.
We begin our searches using the minimal exchange gate sequence ($n=1$).  
If no valid solutions are obtained for a given sequence length, we increment $n$ by 1 and repeat the procedure.
Once an optimal, numerical solution has been obtained, it is sometimes possible to work backwards, to determine the exact rotation angles, as in Figs.~\ref{fig:Bell}, \ref{fig:Ubus2gates}, and \ref{fig:GHZ3Ubus3}.
These identifications can then be confirmed analytically.

\end{appendix}



\begin{thebibliography}{99}

\bibitem{Hanson07}
R. Hanson, L. P. Kouwenhoven, J. R. Petta, S. Tarucha, and L. M. K. Vandersypen, \rmp \textbf{ 79}, 1217 (2007).

\bibitem{Morton11}
J. J. L. Morton, D. R. McCamey, M. A. Eriksson, and S. A. Lyon, Nature \textbf{479}, 345 (2011).

\bibitem{Petta05} 
J. R. Petta, A. C. Johnson, J. M. Taylor, E. A. Laird, A. Yacoby, M. D. Lukin, C. M. Marcus, M. P. Hanson, and A. C. Gossard, Science \textbf{309}, 2180 (2005).

\bibitem{Loss98} 
D. Loss and D. P. DiVincenzo, \pra \textbf{ 57}, 120 (1998).

\bibitem{Levy02} 
J. Levy, \prl \textbf{ 89}, 147902 (2002).

\bibitem{DiVincenzo00}
D. P. DiVincenzo, D. A. Bacon, J. Kempe, G. Burkard, and K. B. Whaley, Nature \textbf{408}, 339 (2000).

\bibitem{Svore05}
K. M. Svore, B. M. Terhal, and D. P. DiVincenzo, \pra \textbf{ 72}, 022317 (2005).

\bibitem{Khaneja02}
N. Khaneja, S. J. Glaser, and R. Brockett, \pra \textbf{ 65}, 032301 (2002).

\bibitem{Oh11} 
S. Oh, L.-A. Wu, Y.-P. Shim, J. Fei, M. Friesen, and X. Hu, \pra \textbf{ 84}, 022330 (2011).

\bibitem{Yung04}
M. -H. Yung, D. W. Leung, and S. Bose, Quant. Inf. Comp. \textbf{4}, 174 (2004).

\bibitem{Benjamin03}
S. C. Benjamin and S. Bose, \prl \textbf{ 90}, 247901 (2003).

\bibitem{Ruderman54} 
M. A. Ruderman and C. Kittel, Phys. Rev. \textbf{96}, 99 (1954); 
T. Kasuya, Prog. Theor. Phys. \textbf{16}, 45 (1956);
K. Yosida, Phys. Rev. \textbf{106}, 893 (1957).

\bibitem{Mizel04} 
A. Mizel and D. A. Lidar, \prl \textbf{ 92}, 077903 (2004).

\bibitem{Scarola05} 
V. W. Scarola and S. Das Sarma, \pra \textbf{ 71}, 032340 (2005).

\bibitem{Hsieh12}
C. -Y. Hsieh, A. Rene, and P. Hawrylak, \prb \textbf{ 86}, 115312 (2012).  

\bibitem{Shulman12}
M. D. Shulman, O. E. Dial, S. P. Harvey, H. Bluhm, V. Umansky, and A. Yacoby, Science \textbf{336}, 202 (2012).

\bibitem{Brunner11}
R. Brunner, Y.-S. Shin, T. Obata, M. Pioro-Ladrie\`{r}e, T. Kubo, K. Yoshida, T. Taniyama, Y. Tokura, and S. Tarucha, \prl \textbf{ 107}, 146801 (2011).

\bibitem{Nowack11}
K. C. Nowack, M. Shafiei, M. Laforest, G. E. D. K. Prawiroatmodjo, L. R. Schreiber, C. Reichl, W. Wegscheider, and L. M. K. Vandersypen, Science \textbf{333}, 1269 (2011).

\bibitem{Gaudreau06} 
L. Gaudreau, S. A. Studenikin, A. S. Sachrajda, P. Zawadzki, A. Kam, J. Lapointe, M. Korkusinski, and P. Hawrylak, \prl \textbf{ 97}, 036807 (2006).

\bibitem{Laird10} 
E. A. Laird, J. M. Taylor, D. P. DiVincenzo, C. M. Marcus, M. P. Hanson, and A. C. Gossard, \prb \textbf{ 82}, 075403 (2010).

\bibitem{Gaudreau11} 
L. Gaudreau, G. Granger, A. Kam, G. C. Aers, S. A. Studenikin, P. Zawadzki, M. Pioro-Ladrière, Z. R. Wasilewski, and A. S. Sachrajda, Nature Phys \textbf{8}, 54, (2011).

\bibitem{Friesen07} 
M. Friesen, A. Biswas, X. Hu, and D. Lidar, \prl \textbf{ 98}, 230503 (2007).

\bibitem{Hill07}
C. D. Hill, \prl \textbf{ 98}, 180501 (2007).

\bibitem{JZhang03pra} 
J. Zhang, J. Vala, S. Sastry, and K. B. Whaley, \pra \textbf{ 67}, 042313 (2003).

\bibitem{Kraus01} 
B. Kraus and J. I. Cirac, \pra \textbf{ 63}, 062309 (2001).

\bibitem{Makhlin02} 
Y. Makhlin, Quant. Info. Proc. \textbf{1}, 243 (2002).

\bibitem{Hill97} 
S. Hill and W.K. Wootters, \prl \textbf{ 78}, 5022 (1997).

\bibitem{Wootters98} 
W.K. Wootters, \prl \textbf{ 80}, 2245 (1998).

\bibitem{Barenco95} 
A. Barenco, C. H. Bennett, R. Cleve, D. P. DiVincenzo, N. Margolus, P. Shor, T. Sleator, J. A. Smolin, and H. Weinfurter, \pra \textbf{ 52}, 3457 (1995).

\bibitem{Nielsen&Chuang} 
M. A. Nielsen and I. L. Chuang, \textit{Quantum Computation and Quantum Information} (Cambridge University Press, Cambridge, UK, 2000).

\bibitem{Vidal04} 
G. Vidal and C. M. Dawson, \pra \textbf{ 69}, 010301 (2004).

\bibitem{Bremner02} 
M. J. Bremner, C. M. Dawson, J. L. Dodd, A. Gilchrist, A. W. Harrow, D. Mortimer, M. A. Nielsen, and T. J. Osborne, \prl \textbf{ 89}, 247902 (2002).

\bibitem{Nielsen99}
M. A. Nielsen, \prl \textbf{ 83}, 436 (1999).

\bibitem{ElzermanAPL04}
J. M. Elzerman, R. Hanson, L. H. Willems van Beveren, L. M. K. Vandersypen, and L. P. Kouwenhoven, Appl. Phys. Lett. \textbf{84}, 4617 (2004).

\bibitem{JZhang04prl} 
J. Zhang, J. Vala, S. Sastry, and K. B. Whaley, \prl \textbf{ 93}, 020502 (2004).

\bibitem{Shim11} 
Y.-P. Shim, S. Oh, X. Hu, and M. Friesen, \prl \textbf{ 106}, 180503 (2011).

\bibitem{Dur00} 
W. D\"ur, G. Vidal, and J. I. Cirac, \pra \textbf{ 62}, 062314 (2000). 

\bibitem{GHZ}
D. M. Greenberger, M. A. Horne, A. Shimony, and A. Zeilinger, Am. J. Phys. \textbf{58}, 1131 (1990).

\bibitem{Mermin} 
N. D. Mermin, \textit{Quantum Computer Science: An Introduction} (Cambridge University Press, Cambridge, UK, 2007).

\bibitem{Galiautdinov08} 
A. Galiautdinov and J. M. Martinis, \pra \textbf{ 78}, 010305(R) (2008).

\bibitem{Briegel01} 
H. J. Briegel and R. Raussendorf, \prl \textbf{ 86}, 910 (2001).

\bibitem{Raussendorf01} 
R. Raussendorf and H. J. Briegel, \prl \textbf{ 86}, 5188 (2001).

\bibitem{Shende09} 
V. V. Shende and I. L. Markov, Quant. Inf. Comp. \textbf{9}, 461 (2009).

\bibitem{DiVincenzo94}
D. P. DiVincenzo and J. Smolin, arXiv:cond-mat/9409111 (1994).

\bibitem{Bodoky07}
F. Bodoky and M. Blaauboer, \pra \textbf{ 76}, 052309 (2007).

\bibitem{Dirac}
P. A. M. Dirac, \textit{The Principles of Quantum Mechanics} (Oxford University Press, USA, 1982).

\bibitem{Feynman} 
R. P. Feynman, \textit{Statistical Mechanics} (Westview Press, Boulder, 1998).

\bibitem{DixonBook}
L. C. W.  Dixon  and  G. P.  Szeg\"{o} (eds.), in \textit{Towards Global Optimisation 2} (North-Holland, Amsterdam, 1978).

\bibitem{HorstBook}
R. Horst and P.M. Pardalos (eds.), in \textit{Handbook of Global Optimization, Vol. 1} (Kluwer Academic, Dordrecht, 1995).

\bibitem{NelderMead}
J. A. Nelder and R. Mead, Comput. J. \textbf{7} 308 (1965).


\end{thebibliography}

\end{document}